\documentclass[conference, 10pt]{IEEEtran}
\IEEEoverridecommandlockouts

\usepackage{amsmath,amssymb,amsfonts}

\usepackage{cite}
\usepackage{url}
\usepackage{algorithmic}
\usepackage{graphicx}
\usepackage{textcomp}
\usepackage{xcolor}

\def\BibTeX{{\rm B\kern-.05em{\sc i\kern-.025em b}\kern-.08em
		T\kern-.1667em\lower.7ex\hbox{E}\kern-.125emX}}

\usepackage{subfig}
\usepackage{fixltx2e}
\usepackage{psfrag}
\usepackage{array}
\usepackage{booktabs}
\usepackage{multirow}
\usepackage{paralist}
\usepackage{mdwmath}
\usepackage{mdwtab}

\usepackage{epstopdf}
\usepackage{pseudocode} 
\usepackage{float}

\usepackage{multicol}

\begin{document}
	
\title{Towards a General Market for Cloud-Edge-IoT Continuum}


\author{Hao Che, Hong Jiang, Zhijun Wang
  \\
  \\
  \texttt{University of Texas at Arlington}\\
  \texttt{\{hche,hong.jiang,zhijun.wang\}@uta.edu}}
  
  
  
  
  
  
  
  
  
  
  
  


\maketitle

\begin{abstract} 
Recent years have witnessed the proposals aiming at enabling Vertical, two-sided markets with a Single Marketplace (or exchange) (VSMs) for computing and data resources/services (products) offerings in a multi-cloud and crowdsourced IoT-edge sensing environment. A VSM is designed vertically from bottom up with a broker being a built-in component of the marketplace.   
While preventing seller lock-in and improving efficiency and availability, a VSM suffers from a key weakness from a buyer's perspective, i.e., the broker and the corresponding marketplace lock-in, which may lead to suboptimal shopping experience for buyers, due to marketplace monopoly by the broker and limited choice of products in the marketplace.  In this position paper, we argue that a Horizontal two-sided market with Multiple Marketplaces (HMM), 
resembling the global stock market, should be developed. In an HMM, different marketplaces may be operated by different parties and sell similar and/or different types of products, e.g., computing and/or sensory data products. A broker is no longer a built-in component of any given marketplace. Instead, it may cover multiple marketplaces at the same time and there can be more than one broker in the HMM. Both the number and types of marketplaces and brokers may grow independently or scale horizontally to meet the growing demand.  A buyer shops for a broker through whom the buyer gains access to the needed products sold in the marketplace(s) the broker covers and from whom the buyer receives various possible services, e.g., discount, value-added, or full services.   
An HMM not only overcomes the key weakness of a VSM but also allows the market to grow incrementally and organically. 
Finally, two example use cases are given to illustrate the benefits of HMM. 

\end{abstract}

\section{Introduction}
\label{sec:intro}
As the cloud, edge and IoT continuum has been evolving into a mature ecosystem, opportunities arise for a user (buyer) to fulfill its computing or (sensory) data analytics job by leveraging the diverse computing and/or data resources/services (or products in short), readily available from various computing and data owners and service providers (sellers) at low cost. Unfortunately, to date, both computing and data products are mostly offered in a one-sided market in which buyers buy products from a single seller, e.g., a cloud service provider. Clearly, to fully utilize the resources/services available in this ecosystem, a two-sided market that admits more than one seller should be in place to allow buyers in the market to shop for and strike deals with potentially more than one seller for the desired products to complete a job and to have the job scheduled and executed using the purchased products. To this end, recent years have witnessed a few proposals aiming at developing Vertical two-sided markets with a Single Marketplace (or exchange) (VSMs) for either computing in a multi-cloud environment (e.g., Mandi \cite{mandi} and Sky Computing \cite{sky}) or data sensing (e.g., mobile-crowdsourcing-as-a-service \cite{crowdpower}) in a crowdsensed cloud-edge-IoT-based environment. 

A VSM is built vertically from bottom up, meaning that the platform that enables a VSM provides all the needed functions for trading in the marketplace, including basic operations, such as bidding/auctioning, market making, (dynamic) product pricing, buyer-and-seller account managing, etc., and the entire lifecycle for trading (e.g., exchanging, clearing and settlement) \cite{thesis} with a broker, a built-in component, serving as the intermediary between buyers and sellers and offering various possible services, ranging from discount bear-bones to rich value-added and full services. 
As a two-sided market, a VSM makes it possible for a buyer to buy products from more than one potential seller and hence, is free from seller or vendor lock-in, e.g., a cloud service provider lock-in \cite{sky}. Moreover, with the help of the broker, 
a buyer has the opportunity to fulfill the job with improved performance and availability at low cost, by exploiting the cost-and-quality tradeoffs of the like products and a wider range of products, as compared with a one-sided market. 
Although promising, a VSM suffers from a key weakness from buyers' perspective, i.e., the broker and the corresponding marketplace lock-in. The implication of this weakness is multifold. Among other things to be discussed in detail in the next section, the broker lock-in may lead to suboptimal shopping experience for a buyer due to marketplace monopoly by the broker and the marketplace lock-in may cause job unfulfillment due to the lack of or unbearably high price tags of the needed products in the marketplace to fulfill the job requirements. 

In this position paper, we argue that a Horizontal two-sided market with Multiple Marketplaces (HMM), resembling the global stock market\footnote{Here, the global stock market refers to a market composed of 
many stock exchanges globally, which can be reached via major brokers, such as Interactive Brokers, Fidelity, and Charles Schwab, who cover both U.S. and foreign market exchanges in 34, 25, and 30 foreign countries, respectively \cite{stock}.}, should be developed. In an HMM, different marketplaces may be owned and operated by different parties and sell similar or different types of products, e.g., computing and/or sensory data products. Meanwhile, more than one broker is allowed in the market, covering one or multiple marketplaces, and multiple brokers may compete with one another in the same marketplace. A buyer shops for a broker through whom the buyer gains access to the needed products sold in the marketplace(s) the broker covers and from whom the buyer receives various possible services. As the demand for the existing or new products grow, both the numbers and types of marketplaces and brokers may grow independently to meet the demand, known as horizontal scaling.    
In other words, an HMM not only overcomes the key weakness of VSMs, but also allows the market to scale horizontally, the key to allow a two-sided market to evolve into a global one covering the entire continuum. 

The rest of the paper is organized as follows. Section 2 gives a brief background introduction of the state-of-the-arts and their limitations. Section 3 makes a case for HMM including the rationales behind HMM and two example use cases that illustrate the benefits of HMM. Finally, Section 4 concludes the paper.

\section{Background and Related Work}

As the cloud, edge and IoT continuum has been evolving into a mature ecosystem, recent years have seen significant development of middleware and orchestration platforms for computing (\cite{santoro2017foggy, nastic2016middleware, saurez2021oneedge, zhang2022ents, zhang2021joint, wojciechowski2021netmarks}) and sensing (\cite{alarbi2018sensing}) over the continuum. 
as well as commercial hybrid cloud platforms \cite{amazon-hybrid, azure-hybrid, google-hybrid} and commercial development platforms for cloud-centric IoT systems
(\cite{amazon-iot, google-iot, azure-iot}). 
They are mostly focused on the technical design aspect, e.g., employing multi-tier, multi-level clustering to improve scalability, much less on the business modeling aspect.
As such, their designs are inevitably underlaid implicitly by certain types of market models, some of which being inherently one-sided. For example, 
some existing platforms require that the resource availability information from the IoT and edge tiers be conveyed to the cloud tier
for centralized control, which may only be viable in a one-sided market in which the sole seller may own and hence, has full control over the resources across all tiers. This may largely limit their applicability in practice, as they may not make business sense in the case where not all the devices in the IoT and edge tiers are owned by the party who also owns the cloud tier.
The work in this paper is to provide our vision as to what business model one should adopt, e.g., who should get involved and who owns what, before a platform is architected. Doing so will ensure that the platform thus architected will indeed make business sense by design.

Relatively fewer proposals on the platform development explicitly address the business modeling aspect, which
to the best of our knowledge, have exclusively based on 
Vertical, two-sided markets\footnote{Readers who are interested in the research work dedicated to the two-sided business modeling (e.g., trading and pricing strategies, and theories) for cloud, edge and/or IoT may refer to \cite{theory, game, auction} and references therein.} with a Single Marketplace or exchange (VSMs)\cite{thesis, mandi, sky, crowdpower} and they can be broadly 
divided into two categories, one for cloud computing with cloud service providers as sellers \cite{thesis, mandi, sky}
and the other for crowdsensing with sensory data collectors as the sellers\footnote{Sensing-as-a-service (SaaS) \cite{sheng2012sensing,perera2014sensing,shi2016edge} or more generally, IoT-based sensing data analytics \cite{analytics1,analytics2} have been gaining popularity in recent years for, e.g., smart city, environmental monitoring, e-Health, digital agriculture, etc. Large bodies of related publications are available. Please refer to recent surveys \cite{sensing1,sensing2,analytics2} and references therein.}.  

In the cloud computing category, an earlier work \cite{thesis} develops a 
single-marketplace-based platform vertically from bottom up focusing on the design of detailed components, in parallel to a 
commodity exchange, including exchanging, clearing, settlement, and rating services, 
with relatively less discussion on the broker role for a sole broker in the marketplace. Another earlier work \cite{mandi} develops, Mandi, a platform-independent framework for multi-cloud computing as a utility in a utility marketplace. Mandi places its focus on the design of various negotiation strategies for trading. Although technically, Mandi allows more than one broker to handle job resource allocation, those brokers are simply functional components of a single brokerage service and hence, from business perspective, Mandi is still a single-broker solution. 
More recently, a group of researchers from UC Berkeley advocates the necessity of a two-sided market for multi-cloud computing, known as Sky Computing \cite{sky}, and subsequently develops a platform, called SkyPilot \cite{skypilot} for trading cloud computing products for batch jobs. SkyPilot enables a marketplace in which a broker assumes significant roles, from basic marketplace operations (e.g., product cataloging and pricing) to product trading, resource allocation and job execution.  

In the crowdsensing category, given that the sensory data must be collected from geo-distributed sensory data owners or collectors through a crowdsourcing mechanism, intuitively and apparently, a market that enables such activities must be two-sided by design, with sensory data consumers as the buyers and sensory data collectors as the sellers. A recent work \cite{crowdpower} on the development of CrowdPower, a platform for mobile crowdsensing-as-a-service is probably the most comprehensive solution in this category to date. Among other things, CrowdPower does detail the roles of a sole broker as an intermediary between sellers and buyers in a single marketplace for real-time sensing. 

In summary, the business models put forward in the above proposals are VSMs by design and hence, suffer from its key weakness, i.e., the broker and the corresponding marketplace lock-in. 

\begin{figure}
     \centering
     \includegraphics[width=1\linewidth]{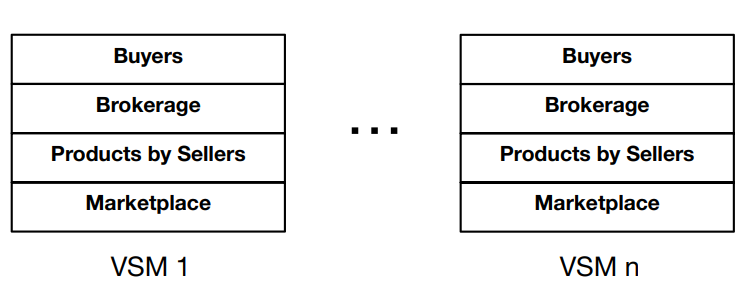}
     \caption{Vertical two-side markets with a Single Marketplace (VSMs)}
     \label{fig:vsms}
 \end{figure}

To further illustrate the implications of the key weakness of VSMs, assume that there are $n$ VSMs in the continuum, as shown in Fig. \ref{fig:vsms}, each being represented by a stack with a lower layer providing services to its immediate upper layer. Specifically, the marketplace at the bottom hosts products from sellers in the marketplace, which in turn are used by the broker to service the buyers on the top. 

First, as one can see, a buyer having a brokerage (or broker) account with the broker in any given VSM can only access the products available in the marketplace in that VSM and hence is locked in with both the broker and the marketplace. 
In this case, what a buyer can buy 
is limited to the available products in that marketplace only, which not only limits the choice of the 
products, but may also render the buyer failing to fulfill the job.  

Using the existing proposed platforms as an example, a buyer in a VSM enabled by CrowdPower \cite{crowdpower} selling sensory data products cannot access cloud computing products in a VSM enabled by SkyPilot \cite{skypilot}. This may make it difficult for the buyer to acquire the needed sensory data and computing products to perform machine-learning model training  using federated learning \cite{fed-service}. This is simply because federated learning may require the buyer to buy sensory data products from sensory data collectors for distributed model training as well as computing products from cloud service providers for the central control of the training process. With the VSM business model, the way to fix this problem is for either of the two VSMs to expand its product line to include the one in the other marketplace, i.e., scaling the VSM marketplace vertically. Such vertical scaling can be daunting, as the types of products sold in the two markets are so different so that the platforms developed for the two are largely incompatible with each other. For instance, while the VSM enabled by SkyPilot mainly sells computing products with relatively stable performance and static pricing structures (e.g., different types of virtual machine instances with different price tags), the VSM enabled by CrowdPower needs to deal with the recruitment of sensory data collectors, who may be participatory or voluntary, static or mobile, data privacy sensitive or insensitive, etc.  

One may still argue that the buyer can get around the above issue by accessing both VSMs via the corresponding brokerage accounts to acquire the needed products to fulfill all the job requirements. This is indeed doable. However, it does not help solve the issues related to the broker lock-in in any of the two VSMs. Moreover, it requires that the buyer manually put together the products bought from different marketplaces to fulfill all the job requirements, which otherwise, could have been done as a value-added service offered by a single broker, if the broker were allowed to cover both marketplaces, as in the case of HMM (see Section \ref{usecase} for an example use case).       

Second, due to the broker lock-in, a buyer in a VSM has no choice but relies on the broker of the VSM to fulfill the job, in the hope that the broker will act in good faith on behalf of the buyer to deliver the service efficiently in terms of both performance and cost. Moreover, the available brokerage services the buyer has access to are limited to those being offered by the broker. For example, the buyer may 
want its model training job to complete with a predefined finishing time and training accuracy, which however, may not be available in 
the value-added service portfolio offered by the broker.

\section{HMM: a Horizontal two-side market with Multiple Marketplaces}
\label{hmm}

\subsection {Why horizontal design?}
The history has taught us that a horizontally designed
system is likely to be more scalable than a vertically designed one. Perhaps the oldest system that managed to reach global scale is the postal service network, which is horizontal by design. It may be viewed as a network of many networks with a logic postal service network
running on top of many transportation networks. The logical link between two adjacent nodes (i.e., the link through which the mails can be delivered from one node, e.g., a distribution center, to the other node) in the postal service network is enabled by one or more (heterogeneous) transportation service networks, e.g., an airline service network and a railway service network. The ability of the postal service network to engage more and more heterogeneous transportation networks to expand the service coverage (i.e., horizontal scaling) is the key to its success of reaching global scale. Now imagine what would happen if the postal service were to be designed vertically from bottom up. One would have to purposely design and implement a transportation network dedicated to the postal service before the postal service itself may be developed and expanding the service coverage requires the expansion of both transportation network 
and postal service network, which can be costly and painstaking and hence, is not scalable. The Internet design closely resembles the postal service design, in the snese that it is also a (IP or layer 3) 
network of many heterogeneous (layer 2) networks \cite{internet}, which can be scaled horizontally. This explains, for a large part, why the Internet also managed to reach global scale. In the same spirit, in what follows, we make a case for HMM, with reference to the global stock market as an analogy.   
\begin{figure}
    \centering
    \includegraphics[width=1\linewidth]{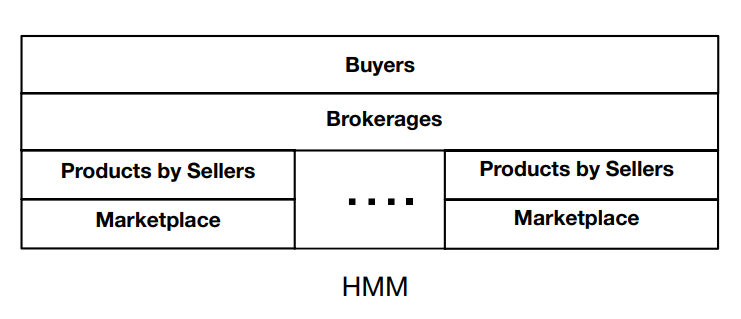}
    \caption{Horizontal two-side market model}
    \label{fig:hmm}
\end{figure}
\subsection {HMM Model}
Analogous to the global stock market involving a set of market exchanges hosting similar or different types of stocks and multiple brokers, each covering one or more market exchanges, the proposed HMM, as shown in Fig. \ref{fig:hmm} in the form of a stack, is composed of a set of marketplaces at the bottom, hosting similar or different types of products offered by the sellers; a set of brokers, each working with one or more marketplaces\footnote{Note that covering more marketplaces does not necessarily mean that a broker will do better to attract customers. Many factors may affect how many and what types of marketplaces a broker may be willing or able to cover, such as operational cost, domain knowledge, available human resources, and business strategies} and offering various possible services; and buyers who may use any of the brokerage accounts to access the products sold in the underlying marketplace(s) and the brokerage services to meet the job requirements for a given job. 

Different marketplaces in an HMM may be owned by different parties, who are responsible for the enabling and normal operations of the corresponding marketplaces, including, but not limit to, auctioning, market-making, product cataloging, account managing for brokers, buyer and sellers, providing and managing trading mechanisms and interfaces. 

A broker, as an intermediary between buyers and sellers, makes use of the trading mechanisms and interfaces provided by the marketplace to (a) provide a customized, user-friendly frontend that enables basic functions for self-directed trading; (b) act on behalf of a buyer to find the needed products from potential sellers to fulfill the job requirements; and (c) with the purchased products, act on behalf of a buyer to perform job resource allocation, scheduling and execution. A broker who provides only service (a) may be considered as a discount broker, whereas one who offers services (a), (b) and (c) may be viewed as a full-service broker. A broker of any type may also provide a portfolio of value-added services. For example, a discount broker may offer, as a value-added service, tables of compatibility sets of products \cite{sky, skypilot} and an online computer-aided tool that can assist a buyer to analyze and compare the pros and cons and cost features of the products in each compatibility set. As another example, a full-service broker may offer job DAG-workflow performance guaranteed value-added services.   

Note that the above role assignment for a marketplace versus a broker is by no means set in stone, and much more in-depth study in the future as to what exact roles each need to play is warranted. Instead, it is meant to emphasize the importance of drawing a clean boundary between the roles played by a marketplace and a broker in HMM. The objective is to limit as much as possible the coupling between the brokerage layer and the marketplace layer to allow ease of horizontal scaling, i.e., to allow the two layers to scale independent of one another. For example, an HMM initially may only run a single marketplace selling computing products from cloud service providers. As federated learning for machine learning model training \cite{fed-service} becomes increasingly popular, the HMM may decide to expand its market by engaging one more marketplace selling sensory data products. This, in turn, attracts more buyers to the HMM and hence, increases the demand for brokerage services. As a result, 
a broker who initially only covers the marketplace for computing products may decide to also cover the other marketplace. By doing so, the broker can offer federated learning as a value-added service \cite{fed-service} for its customers using the products from both marketplaces. Also more brokers may join the HMM, as the demand for brokerage services further increases. This example clearly demonstrates that as the customer demand for certain types of products increases, it automatically triggers the growth of the number and types of marketplaces, which in turn, triggers the growth of the brokerage services. This auto-scaling or organic growing capability of the HMM in response to the demand increase is made possible only if the two layers are loosely coupled (note that complete decoupling of the two layers is impractical, as a broker must adapt to the specific trading mechanisms and interfaces offered by individual marketplaces).  

\subsection{Relation to the existing VSMs}
One may argue that the reason why the existing two-sided market with a Single Marketplace (SM) proposals only consider a single marketplace is not because the authors are unaware of the possibility of horizontal scaling, but because the development of two-sided-market-based platforms over the continuum is still in its infancy and hence, when designing such a platform, it suffices to consider only one marketplace to start with. We agree that this may well be the case and we also anticipate that it may be implicitly understood by many that HMM would be the result of natural progression of SMs, as such markets proliferate over the continuum. 

Assuming that the above understanding is correct, we hereby acknowledge that the use of adjective, "vertical", to label all the existing SMs may not have done justice to those SMs with no vertical scaling in mind. Nevertheless, we believe that this position paper does make an important contribution. Namely, it is the first paper that explicitly articulates the need for and the importance of the HMM business modeling to allow two-sided markets to grow into one that may reach global scale, and in relation to the design of an SM, the importance of weakening the coupling between the broker and the marketplace of the SM in terms of both marketplace ownership and operational functions. 
Putting it in another way, a key take away of this position paper is that in developing an SM, it is crucial to clearly demarcate the boundary between the broker and the marketplace so that the broker can plug its brokerage services in and out of the marketplace without affecting the normal operation of the marketplace, making it easy for the SM to join other SMs to form a single HMM, as SMs proliferate.      

\subsection{Things not considered in this paper}
This position paper solely focuses on the business modeling, or more precisely, the business organization aspect of a complete platform solution for the effective utilization of the computing and data products available in the cloud-edge-IoT continuum to best benefit product consumers and producers. 
As mentioned in Section 2, there are significant research proposals, middleware, orchestration and development platforms available, which collectively have addressed various system design challenges not being touched upon in this position paper, e.g., multi-tiered design for scalability \cite{mortazavi2020feather}, programming frameworks for serverless computing \cite{hybrid}, collaborative sensing for stable sensing quality in  mobile sensing environments \cite{collaborative-sensing,virtual-sensor}, limiting data exposure for privacy and security \cite{ding2022privacy,garrido2022revealing},  just to name a few. 

The main objective of this work is to identify HMM as a business model that is likely to thrive and grow into one that covers the entire continuum globally. As such, this position paper has mainly focused on the high-level description of the HMM stack and the demarcation between the brokerage layer and the marketplace layer. The objective is to stimulate discussions and debates among researchers and practitioners in related research communities and industries. 
The detailed designs of the building blocks of HMM are left to be studied in the future. To this end, one may borrow and customize the ideas and solutions from the existing designs, such as crowd recruitment \cite{crowdpower}, negotiation \cite{mandi}, trading and pricing \cite{theory, game, auction} strategies, product cataloging, job resource allocation and scheduling \cite{skypilot}, and other essential operational functions for a marketplace \cite{thesis}. On top of these basic building blocks of HMM, the future research must also address many other design challenges/issues not accounted for in this paper. One example is how to put together the products purchased from different marketplaces to enable job execution with desired performance and/or cost saving. This could be a challenge to be tackled by either a buyer who uses a discount brokerage service or a broker who offers it as value-added services to the buyer.  
Tackling this challenge is an interesting research topic in its own right, noting that a big part of the broker design for SkyPilot \cite{skypilot} evolves around addressing a similar challenge, i.e., how to coordinate the job execution using the computing resources purchased from different cloud service providers. 



\section {HMM Use Cases}
\label{usecase}
\begin{figure}
    \centering
    \includegraphics[width=1\linewidth]{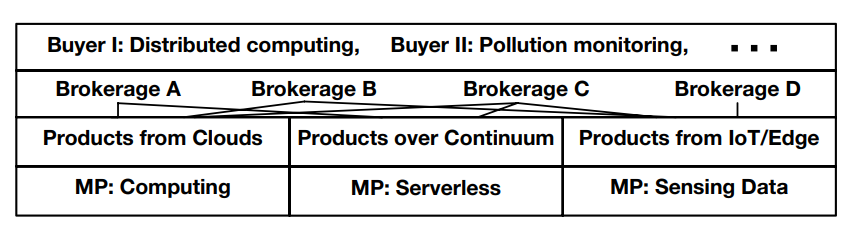}
    \caption{Two HMM use cases}
    \label{fig:use-case}
\end{figure}

In this section, we use two use cases to demonstrate the benefits of HMM.  Fig. \ref{fig:use-case} depicts a hypothetical HMM. In this market, there are three marketplaces (MPs), i.e., MP-1, MP-2, and MP-3, selling computing products from cloud service providers, serverless products across the continuum from serverless service providers, and sensory data products from crowds, respectively, and four brokers, i.e., brokerages A, B, C, and D, covering MP-1 and MP-2, MP-1 and MP-3, all three marketplaces, and MP-3, respectively. 

We consider two use cases: Buyer I who needs to buy products to fulfill a distributed, big-data computing job; and Buyer II who needs to set up an environment (e.g., pollution or flooding) monitoring and alert system based on crowdsensing in a city using a machine learning algorithm for data analytics and federated learning for model training \cite{analytics1, analytics2}. 

\subsection{Use Case I}
For this use case, Buyer I may use products offered by either MP-1 or MP-2, depending on the characteristics of the job. For example, if it is a one-time job or a recurring job with long recurring intervals, Buyer I may choose MP-2 that offers serverless products to minimize the cost, thanks to the pay-as-you-go nature of the serverless products. On the other hand, if the job is recurring with high recurring frequency and lasts for a sustained period of time, Buyer I may prefer MP-1 over MP-2 because in this case, it may be more cost effective to buy and hold the computing resources for the entire job duration. 

In the case where Buyer I needs to use serverless products from MP-2, Buyer I has the choice of using either brokerages A or C, as they both cover MP-2. Depending on what services and service fees the two brokerages have to offer, Buyer I may choose the one that  meets its overall job requirement at lower cost. For example, if Buyer I has the know-how as to how to decompose the response time target for the job DAG workflow into individual query delay budgets for queries corresponding to individual stages in the DAG workflow, Buyer I may want to use a low-cost value-added service that provides per query delay target guarantee to meet its job response time target. Otherwise, Buyer I may need to use a high-cost value-added service that provides per job response time guarantee to meet the job response time target. In either case, Buyer I will have to make a decision as to which brokerage service should be used, based on the actual value-added services offered by the two brokerages and the associated fees. 

In the case where Buyer I needs to use products offered by MP-1, Buyer I has the choice of using brokerage, A, B or C, as all three cover MP-1. Again, depending on how knowledgeable the buyer is with regard to the resource allocation and management, as well as job scheduling, and what services and service fees the three brokerages have to offer, Buyer I will select the one that meets its overall requirement the best. 

The above use case clearly demonstrates the flexibility of HMM to allow a buyer to choose the right marketplace(s) to buy the needed products as well as the right broker with the desired brokerage service(s). 

\subsection{Use Case II}
Assuming that Buyer II is a city council who has in-house expertise and general know-how for setting up, operating and managing a distributed system. With limited budget allocated, the city council decides to use the low cost products and services as much as possible. This use case involves two different jobs: setting up and running the monitoring and alert system; and performing (periodic) federated-learning based model training. 

To fulfill the first job, Buyer II needs to buy computing products from MP-1 to serve as the master (i.e., the frontend and central control of the system) in cloud and sensory data products sold by collectors in the same city as Buyer II from MP-3. To this end, Buyer II may decide to use a lowest-cost discount service offered by one of the two brokerages, B and C, who cover both marketplaces, and shop for the best computing and sensory data products in terms of cost-quality tradoffs to set up the monitoring system all by itself. 
Alternatively, assuming that the in-house engineers are not familiar with the recruitment and negotiation strategies involved in the process of purchasing sensory data products and by checking out the available value-added services in the portfolios offered by the two brokerages, Buyer II may come to the conclusion that it is in fact more cost-effective to use a desired environment monitoring and alert SaaS (i.e., Sensing-as-a-Service) \cite{shi2016edge}, a turn-key value-added service offered by one of the brokerages, say, brokerage B, for the initial set up of the monitoring system\footnote{Note that both brokerages B and C are capable of offering SaaS, since both have access to the needed products, i.e., computing from MP-1 and sensory data from MP-3, to enable such services.}. After the system is up and running, the in-house engineers then take over the control of the entire system to keep the cost low. 

To fulfill the second job, if the budget allocated for model training is tight, Buyer II may perform federated-learning-based model training using the sensory data already collected by the collectors, while running the system. But if the budget allows, via either brokerage A or C, Buyer II may also use the serverless federated learning products \cite{fedless} offered by MP-2 for model training with the sensory data collected in some other cities for broader training data samples and hence, better training. 

The above use case reveals the ability of HMM to allow a buyer to fully explore the various products offered in different marketplaces and diverse brokerage services offered by brokers. 

In summary, the above two use cases clearly demonstrates that HMM indeed has the potential to offer a multitude of benefits in terms of scalability, diversity, availability, flexibility and extensibility.


\section{Conclusions}
In this position paper, we advocate and make a case for the development of a general market, 
HMM, for the cloud, edge and IoT continuum. In an HMM, different marketplaces may be operated by different parties and sell similar and/or different types of products. As an intermediary between buyers and sellers, a broker may cover multiple marketplaces at the same time and there can be more than one broker in the HMM. Both the number and types of marketplaces and brokers may grow independently or scale horizontally to meet the growing product demand. A buyer shops for a broker through whom the buyer gains access to the needed products sold in the marketplace(s) the broker covers and from whom the buyer receives various possible services, e.g., discount, value-added, or full services.   
An HMM not only overcomes a key weakness of the traditional vertical two-sided markets with a single marketplace --- the broker and the corresponding marketplace lock-in, but also allow the market to grow incrementally and organically. 

\bibliographystyle{abbrv}
\bibliography{DSSP-ref}

\end{document}